\documentclass[format=sigconf]{acmart}
\renewcommand\footnotetextcopyrightpermission[1]{}
\AtBeginDocument{%
  \providecommand\BibTeX{{%
    \normalfont B\kern-0.5em{\scshape i\kern-0.25em b}\kern-0.8em\TeX}}}

\usepackage{float}
\usepackage{url}
\usepackage{hyperref}

\begin{document}

\title{Timestamp-independent Haptic-Visual Synchronization}

\author{Yiwen Xu}
\email{xu_yiwen@fzu.edu.cn}
\affiliation{%
	\institution{Fuzhou University}
	\country{ }}
\author{Liangtao Huang}
\email{211120101@fzu.edu.cn}
\affiliation{%
	\institution{Fuzhou University}
	\country{ }}
\author{Tiesong Zhao}
\email{t.zhao@fzu.edu.cn}
\affiliation{%
	\institution{Fuzhou University}
	\country{ }}
\author{Liqun Lin}
\email{lin_liqun@fzu.edu.cn}
\affiliation{%
	\institution{Fuzhou University}
	\country{ }}
\author{Ying Fang}
\email{ fy@fzu.edu.cn}
\affiliation{%
	\institution{Fuzhou University}
	\country{ }}
\begin{abstract}
 The booming haptic data significantly improves the users' immersion during multimedia interaction. As a result, the study of Haptic, Audio-Visual Environment (HAVE) has attracted attentions of multimedia community. To realize such a system, a challenging task is the synchronization of multiple sensorial signals that is critical to user experience. Despite of audio-visual synchronization efforts, there is still a lack of haptic-aware multimedia synchronization model. In this work, we propose a timestamp-independent synchronization for haptic-visual signal transmission. First, we exploit the sequential correlations during delivery and playback of a haptic-visual communication system. Second, we develop a key sample extraction of haptic signals based on the force feedback characteristics, and a key frame extraction of visual signals based on deep object detection. Third, we combine the key samples and frames to synchronize the corresponding haptic-visual signals. Without timestamps in signal flow, the proposed method is still effective and more robust to complicated network conditions. Subjective evaluation also shows a significant improvement of user experience with the proposed method.
\end{abstract}



\maketitle
\section{Introduction}
The recent developments of multimedia technology also cultivate the user requirements on multimedia contents that a more immersive multimedia system is imperative.  As an emerging multimedia signal, the haptics provides newfangled and authentic user perception beyond the reigning audio-visual  signals. Thus, the Haptic, Audio-Visual Environment (HAVE) has raised concerns of researchers with widespread applications in remote surgery, remote training/education, etc~\cite{7593456, 8470161, 9372659, 8399482}.
\begin{figure}[htbp]
	\centering
	\includegraphics[width=\linewidth]{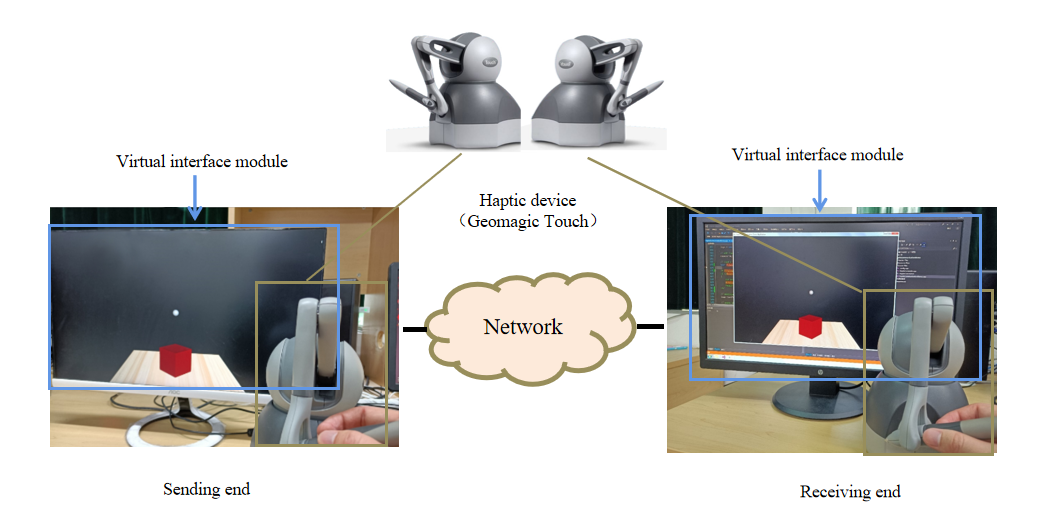}
	\caption{Our simulation platform for haptic-visual signal delivery}
	\Description{Our simulation platform for haptic-visual signal delivery}
	\label{picture1}
\end{figure}

Similar to conventional audio-visual signals, the haptic signal can also be affected during network fluctuations or congestions. In multimedia case, the haptic signal may lose synchronization with other signals, e.g. images, videos. As reported, the haptic-visual asynchronization greatly influences the user experience.  Qi et al.~\cite{6664891} implemented several experiments to explore the impact of the delay between video and haptic signals on the quality of users' experience. The results showed that all the Mean Opinion Score (MOS) values decreased with the inter-flow synchronization error.  The works from Aung et al.~\cite{9293934} also confirmed the above conclusion.

To address this issue, the haptic-visual synchronization is desired. The system examines the synchronization status of signals in real time and adjusts the corresponding signals immediately when an asynchronization is found. To our best knowledge, there is still a lack of synchronization method that is designed for haptic-visual signals, although we have seen enormous models for audio-visual asynchronization detection. In the state-of-the-art HAVE systems, the haptic-visual synchronization is achieved by the timestamp method~\cite{ca667f9f384e4eb8aa68f17b56ea4f67, 4379554} that was designed for generic signals. The timestamp-dependent method embeds the timestamps in signal stream to avoid synchronization drift. The receiving-end detects the signal synchronization status based on the timestamps and the system clock. However, the timestamp-dependent method has its drawbacks. Firstly, in sending end, the timestamps are usually added after frame synchronization, format conversion or pre-processing, where the delay derived from these operations are not compensated~\cite{ca667f9f384e4eb8aa68f17b56ea4f67}. Thus this signal asynchronization in sending end will take to and always exist in the receiving end. Secondly, as the sending and receiving ends have different system clocks (the same frequency), the initial delay and frequency offset caused by dynamic environments also  lead to signal asynchronization.

In this paper, we propose a first-of-its-kind timestamp-independent synchronization method for haptic-visual signals. There exists a strong sequential correlation between haptic-visual signals, which can help the judgment of signal synchronization state without crystal oscillators or timestamps. Inspired by this, we propose a timestamp-independent haptic-visual synchronization model that successfully detect and eliminate asynchronization phenomena in our simulation platform in Fig. 1. Our contributions are summarized as follows.

\textit{The sequential correlation between haptic-visual signals.} We build a multimedia communication platform with both haptic and visual signals, based on which we observe a strong correlation between the two signals during haptic-aware interaction. This intrinsic correlation is further utilized to design our synchronization model.

\textit{The key sample/frame extraction during haptic-visual interaction. }We exploit the statistical features of haptic-visual signals and then develop learning-based methods to extract key samples and key frames in haptic and visual signals, respectively.

\textit{The asynchronization detection and removal strategy.} Combing the correlation and key samples/frames, we are able to detect and eliminate asynchronization when the registration delay is larger than a threshold. Experimental results with subjective evaluations validate the effectiveness of our method.


\section{RELATED WORK}

\textbf{ Haptic-based interaction.} Introducing haptics into the process of human-computer interaction can effectively improve the immersion and realism of the multimedia system, hence, researchers have been conducting research on Haptic-based Interaction System (HIS) since the 1990s ~\cite{2019Error}. The work in~\cite{9034103} proposed a typical HIS framework, which consists of a master control end, a network, and a controlled end. The master control end usually consists of an haptic operator and a human-computer interaction module, which implements the acquisition and pre-processing of haptic signal and transmits them to the controlled end via the network. In the controlled end, a remotely controlled robot or a controlled operator executes the remote interaction commands from the main control end, and at the same time feeds the scene information to the main control end. Our haptic-based simulation platform presented in Section 3 is also according to the above framework.

HIS has been used in a variety of applications. Ilaria et al.~\cite{2018Wearable} designed an immersive haptic VR system for rehabilitation training of children with motor neurological disorders, which significantly improved the effect of rehabilitation training. Zhou et al.~\cite{8616509} proposed an approach with visual and haptic signals, which help physicians to perform surgeries accurately and effectively and furthermore reduce their physical and cognitive burden during surgery. Chen et al.~\cite{2019Application} designed a remote training system with force feedback for power grid operation training. It avoided the collision between the manipulator and steel bars, which helps guide operators to reduce operational errors and complete tasks efficiently. Varun et al.~\cite{8798006} also introduced haptics into a VR-based training system to enhance training immersion, effectiveness and efficiency. The use of HIS for online shopping~\cite{BD2019A,2019Deformation} can improve the realism of the shopping experience and help visually impaired patients enjoy the convenience of online shopping. HIS can also be used in outdoor search and rescue scenarios to avoid collisions by providing tactile guidance~\cite{8007258}. In industry, HIS is usually used to enhance the operation ability of robots. For example, the work in~\cite{2019In} equiped robot with bionic haptic manipulators to help it having more stable grasping ability in teleoperation tasks. In ~\cite{2019Intuitive}, the operator controls the robot to perform teleoperation in real time by means of a pneumatic haptic feedback glove. Apparently, the HIS is widely used and worthy of further investigation.

\textbf{Haptic-based transmission and synchronization.} Haptic interaction enables real-time perception, manipulation and control of real or virtual objects. Compared to video, audio, or image, the transmission of haptics is more tolerant of data loss and bandwidth but has higher requirements for the latency between signals. To ensure more natural interactive operations, haptic-based multimedia signal transmission requires better inter-signal synchronization. However, to the best of our knowledge, current research on synchronization of visual-haptic signals is mainly focused on studying the impact of visual-haptic asynchronization on user experience, while few research has been conducted on synchronization detection and adjustment of visual-haptic signals, and there is still room for improvement in this area.

The researches on synchronization algorithms for audio-visual signals can be used as good references for the research on visual-haptic signals. Timestamp-dependent synchronization method is the most commonly used audio-visual synchronization method (and the method is currently adopted by the video-haptic system), but the method has some shortcomings (summarized in Section 1). To solve these shortcomings, researchers have proposed some improvement algorithms. For example, the works in~\cite{20173D, 8461853} utilized the correlation between audio-visual signals for synchronization detection. They extract lip pictures in video frames and then compare them with the features of audio signal through a deep-learning-based model to determine the synchronization status of audio-visual signals. The limitation of this method is that the video frame must contain the lip region. Yang et al.~\cite{4284763} proposed a watermark-based method to keep the synchronization of audio-visual signal. But this method have a disadvantage that the "watermark" is not well adapted to the video or audio signal when applying conversion, aspect ratio conversion or audio downmixing~\cite{2012Assessing}. In the work of~\cite{4607743}, the features of the audio and visual signals are converted into robust hash codes. At the receiving end, the received signal is used to generate the "signature" again and compare it with the received "signature" to obtain the inter-stream delay of the audio-visual signal.

The asynchronization may occur during the transmission and decoding recovery of multimedia signals due to various factors such as network environment and clock drift. Obviously, signal asynchronization can greatly affect the quality of user experience, so when it occurs, an removal algorithm should be adopted to remove the asynchronization. In this work, we have addressed both asynchronization detection and removal of haptic-visual signals with deep learning. The detailed motivation and explanation of our method is discussed as follows.

\section{Haptic-Visual Correlations}
To design and test our haptic-visual synchronization method, we build a simulation platform with haptic-visual signal delivery. As shown in Fig.~\ref{picture1}, we use the virtual interaction module to design a haptic-visual interaction scenario, where a human user manipulates a virtual ball to push a virtual box. A Geomagic Touch is deployed to connect the real and virtual world: on one hand, it sends the human instructions to the virtual ball; on the other hand, it collects the force feedback of the virtual ball and sends the corresponding signals back to the human user. This haptic interaction is achieved with the kinesthetic signal, which is a major component of haptic information.


Besides of the haptic signals captured by Geomagic Touch, the sending-end also records the visual contents of virtual space, resulting in a high-definition video at a resolution of 1920 $\times$ 1080.  Then the video is compressed by High Efficiency Video Coding (HEVC) and subsequently delivered with haptic signal by network via User Datagram Protocol (UDP). Finally, the receiving-end combines both haptic and visual signals for a more immersive telepresence, where another user can watch the scene in real time and also feels the haptic sensing via a haptic device.

\begin{figure}[htbp]
	\centering
	\includegraphics[width=\linewidth]{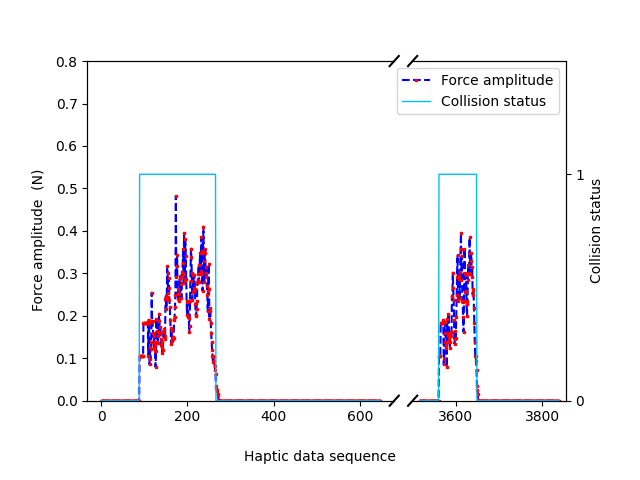}
	\caption{An example of haptic-visual correlations}
	\Description{An example of haptic-visual correlations}
	\label{picture2}
\end{figure}

The haptic and visual signals should be fully synchronized under normal conditions. Based on this simulation platform, we can observe the sequential correlation between haptic and visual signals. As shown in Fig.~\ref{picture2}, there exists strong haptic signal fluctuations when the virtual hand (i.e. the ball) is on collision with another object. When the virtual hand visually touches the box, the force amplitude of haptic changes simultaneously. As the two objects are closer, the force amplitude is also higher; and vice versa. The force amplitude recovers to a constant when all objects are detached. These changes are also intuitive to the human users when operating a haptic-aware handle.

This intrinsic correlation inspires us to design a synchronization strategy. A sharp increase of force amplitude indicates a collision between the virtual hand and another object, while a sharp decrease implies a detachment between objects. If these deductions are inconsistent with the machine vision, we can conclude that there exists an asynchronization between haptic and visual signals and thus change the signal flows.

\begin{figure*}[htbp]
	\includegraphics[width=0.85\textwidth,height=0.34\textwidth]{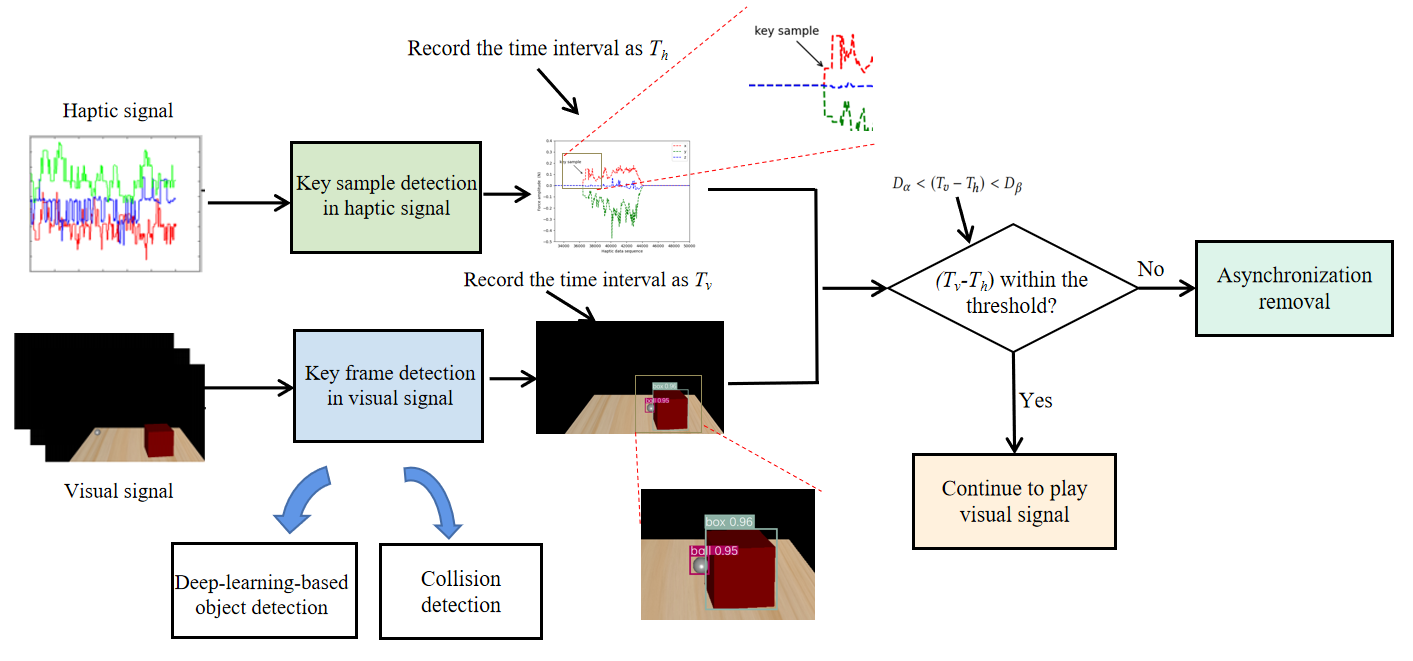}
	\caption{The flowchart of our proposed method}
	\Description{picture3}
	\label{picture3}
\end{figure*}

\section{Proposed Method}

Based on the above analysis, we propose the timestamp-independent synchronization method as shown in Fig.~\ref{picture3}. Firstly, we extract the key samples in haptic signal where the amplitude is intensively increased from near zero. Secondly, we extract the key frames in visual signal where the visual collision happens. Thirdly, we compare the time intervals of these key samples/frames to detect asynchronization phenomena. If a pair of time intervals (namely $T_h$ and $T_v$) are with a large difference, the haptic-visual asynchronization is found and further fixed. Note here the object collision frequencies are low in real world, therefore, we can easily identify different pairs of time intervals. In the following subsections, the key sample detection, key frame detection, threshold selection, asynchronization removal and the overall method are elaborated, respectively.

\subsection{Key sample detection in haptic signal}
For haptic signal, the key samples are easily obtained for it consists of three one-dimensional signals (in x-axis, y-axis and z-axis). A sharp increase of force amplitude is found when its difference in any dimension is larger than a threshold (namely $F_{th}$). In this work, $F_{th}$ is empirically set as 0.05. In additions, to reduce the effects of noise, we employ a Gaussian filter with kernel size 5 as a pre-processing of haptic signal.

\begin{figure}[h]
	\centering
	\includegraphics[width=\linewidth]{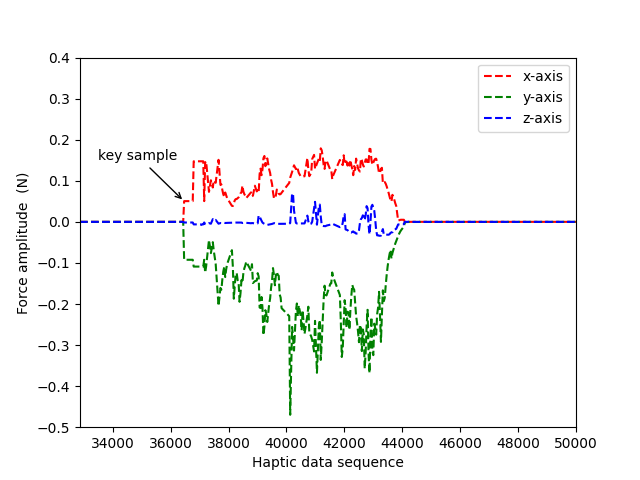}
	\caption{An example of key sample detection}
	\Description{An example of key sample detection}
	\label{Figure_4}
\end{figure}

An example of this step is shown in Fig.~\ref{Figure_4}. An operation with force signals in three dimensions is presented, where all sharp increases are successfully detected and labeled as key samples. Correspondingly, their time intervals (i.e. $T_h$) are recorded for further comparison.

\subsection{Key frame detection in visual signal}
The objective of key frame detection is to find the time intervals when the virtual hand touches the box. Essentially, it consists of two modules: object detection and collision detection. The first module identifies all objects, while the second module determine whether object collision occurs. Both modules are achieved by computer vision methods.
\subsubsection{Object detection:}We select the popular YOLO network to recognize all objects in the video. With deep network, the YOLO network extracts the deep features of different objects and scenarios, thereby achieving object recognition with high accuracy. In particular, we employ the V3 of YOLO network~\cite{8886369} by jointly considering the efficiency and complexity.

We established our image database for training the YOLO V3 network. We acquired 1000 images from visual signals with an image size of 1600×900 pixels. Then the images were labeled via a label making tool (the application software of labelImg). We used rectangle to bound the balls in the images and labeled them as "ball", and accordingly, bound the boxes and labeled them as "box". All the labels were saved with xml files for using during training. The 800 images in this database are employed as the training set and the other 200 images are the test set.

The loss function plays an important role in the YOLO network. In this work, the position information of the ball and box are the target of the network. Therefore, the target's  error of center coordinate in the form of squared difference is first taken into account in the loss function; then, in order to obtain the accurate bounding rectangle, the wide and high coordinate error in the form of cross-entropy is utilized; finally, as the detection of multiple categories of targets (ball and box) are involved, the category error in the form of cross-entropy must be considered. Hence, the loss function used in this work is:
\begin{equation}
	\begin{aligned}
			Loss = \lambda_{\textit{coord}} {\sum_{\textit{}i=0}^{S^{2}}} {\sum_{\textit{j }=0}^{B}} {{\textit{I}}^{\textit{obj}}_\textit{{ij}}} {[{{(\textit{x}_{\textit{i}}-\textit{\^{x}}_i)}}}^{2}+{{{(\textit{y}_{\textit{i}}-\textit{\^{y}}_i)}^{2}]}} +\\
			\lambda_{\textit{coord}} {\sum_{\textit{}i=0}^{S^{2}}} {\sum_{\textit{j }=0}^{B}} {{\textit{I}}^{\textit{obj}}_\textit{{ij}}} {[{{(\sqrt{w_i}-\sqrt{\hat{w}_i})}^{2}}}+{{{(\sqrt{h_i}-\sqrt{\textit{\^{h}}_i})^{2}}]}}- \\
			{\sum_{i=0}^{S^{2}}} {\sum_{\textit{j }=0}^{B}} {{\textit{I}}^{\textit{obj}}_\textit{{ij}}}
			{[\hat{C}_i{\log(C_i)}+{(1-\hat{C}_i)}{\log(1-C_i})] }-\\
			{\lambda_{noobj}}{\sum_{i=0}^{S^{2}}} {\sum_{\textit{j }=0}^{B}} {{\textit{I}}^{\textit{noobj}}_\textit{{ij}}}
			{[\hat{C}_i{\log(C_i)}+{(1-\hat{C}_i)}{\log(1-C_i})] }-\\
			{\sum_{i=0}^{S^{2}}} {I_{ij}^{\textit{obj}}}{\sum_{c\in{classes}}}{[\hat{P}_i{\log(P_i)}+{(1-\hat{P}_i)}{\log(1-P_i})]
				}	
		\end{aligned}
\end{equation}

\noindent where the first row indicates the error of the center coordinates, \textit{S} represents the grid size, \textit{B} represents the bounding rectangle. $I_{ij}^{obj}$ denotes whether  targets are  in the rectangle, and its value is 1 if there is a target in the bounding rectangle at grid (\textit{i}, \textit{j}), and 0 vice versa. $x_i$ and $y_i$ represent the true center coordinates, $\hat{x}_i$ and $\hat{y}_i$ represent the predicted center coordinates. The second row represents the error of the width and height of the predicted rectangle, in which $w_i$ and $h_i$ represent the true width and height, $\hat{w}_i$ and $\hat{h}_i$ represent the predicted width and height. The third and fourth rows indicate the error of the confidence level, where $C_i$ denotes the true confidence level, and $\hat{C}_i$ denotes the predicted confidence level. The fifth row denotes the error of classification, where $P_i$ and $\hat{P}_i$ denote the true and the predicted categories, respectively. $\lambda_{coord}$ and $\lambda_{noobj}$ are the weights which will be trained as  hyperparameters of network.

 \begin{table}[t]
	\caption{The hyperparameter settings in model training}
	\label{tab1+1}
	\begin{tabular}{cccccc} 
		\toprule
epoch&batchsize&$\lambda_{\textit{coord}}$&
$\lambda_{\textit{noobj}}$&learning rate\\
		\midrule
300	&16	&0.5&	0.5	&cosine decay\\
		\bottomrule
	\end{tabular}
\end{table}

The main hyperparameters used in training are set as shown in Table~\ref{tab1+1}. Among them, the learning rate is set as cosine decay as follows:
\begin{equation}
	lr =[1+cos(x\times\dfrac{\pi}{2\times epoch})] \times 0.95+0.05.
\end{equation}
With this method, the training module has larger learning rate at the beginning to accelerate the training speed, and then the learning rate decreases with the increasing number of training epochs to more easily find the optimal solution. Experimental results show with our training strategy, the loss function converges to a very low level and is almost stable after 300 epochs of training.

\begin{figure}[b]
	\centering
	\includegraphics[width=\linewidth]{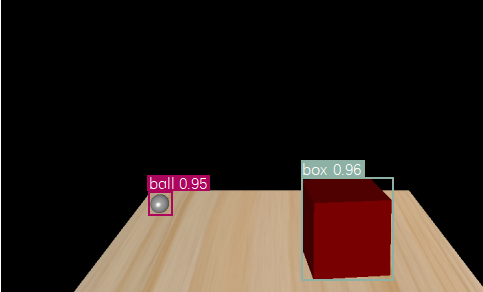}
	\caption{An example of object detection}
	\Description{An example of object detection}
	\label{picture4}
\end{figure}

After training, an example of recognition result is shown in Fig.~\ref{picture4}, in which the virtual hand (i.e. the ball) and the box are detected, with their borders labeled by rectangular frames.

\subsubsection{Collision detection:} We determine whether a collision happens based on the aforementioned rectangular frames. Let $(X_1, Y_1)$ and $(X_2, Y_2)$ denote the top-left locations of virtual hand (i.e. the ball) and any object as target in 2D space, and $(H_1, W_1)$ and $(H_2, W_2)$ denote the sizes of corresponding rectangular frames, the condition of no collision is:
\begin{equation}
	(Y_1+H_1>Y_2)||(X_1+W_1<X_2)||(Y_1<Y_2+H_2)||(X_1>X_2+W_2).
	\label{eq1}
\end{equation}

\noindent Otherwise, the collision of objects is found. At the time of collision found, we extract the corresponding video frame as the key frame of visual signal and record the time interval as $T_v$, which is further utilized for asynchronization detection.

\subsection{The synchronization threshold}
During haptic-visual delivery and playback, we can easily identify each key sample/frame pair considering the corresponding time intervals are usually very close to each other. For a pair of time intervals $T_h$ and $T_v$, their difference is set as a criterion of haptic-visual asynchronization. A synchronization of signals is guaranteed if:
\begin{equation}
	D_\alpha <(T_v-T_h) < D_\beta,
	\label{eq2}
\end{equation}
\noindent where $D_\alpha$ and $D_\beta$ refer the lower and upper bound of perception threshold.

\begin{figure}[b]
	\centering
	\includegraphics[width=\linewidth]{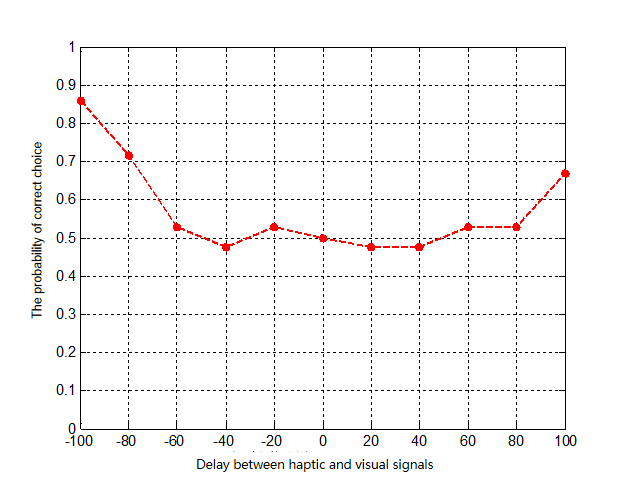}
	\caption{Subjective results of synchronization threshold}
	\Description{Subjective results of synchronization threshold}
	\label{picture5}
\end{figure}

As results from subjective test can be more consistent with users' perception experience, we design an subjective test to determine $D_\alpha$ and $D_\beta$. Our test strictly follows the subjective test manual ITU-R BT.500~\cite{123} with the following steps. First of all, we recruit 21 testees without prior knowledge of haptic coding or delivery. Then, we use the 2-alternative force choice method to perform the test. Each session of test consists of two randomly presented haptic-visual segments: with and without delay. The delay can be negative or positive with a range from -100ms to 100ms with an interval of 20ms. Each testee is asked to choose one segment that he/she cannot feel delay between the two. Finally, for each session, the probability of correct choices, which is obtained by Eq.~(\ref{eq3}), is recorded.
\begin{equation}
	p_i=\dfrac{n_i}{N},
	\label{eq3}
\end{equation}
\noindent where $n_i$ denotes the number of testees who are make a correct choice in the $i$-th delay. $N$ denotes the total number of testees.

As shown in Fig.~\ref{picture5}, the probability of correct choices is around 0.5 when the delay of visual signals ranges from -60ms to 80ms. In other words, the human users cannot perceive the difference between delayed and non-delayed signals in this range. Therefore, we set the threshold of synchronization as $D_\alpha=-60ms, D_\beta=80ms$.

\subsection{Asynchronization removal}

To adjust the signal stream and remove asynchronization phenomena, a general method is to select a main stream and set the remaining as auxiliary streams. When asynchronization occurs, all auxiliary streams are adjusted to be synchronized with the main stream. As reported in~\cite{YutakaIshibashi2018Media}, the human perception of haptic signals is very sensitive that only the haptic signals above 1kHz provide smooth experience to users. This frequency is significantly higher than visual signals. Based on this fact, we utilize the haptic signal and visual signal as the main stream and the auxiliary stream, respectively. For synchronization, the visual signal is moved to be consistent with the haptic signal.

In multimedia communication system, the receiving-end usually set a buffer zone to cache all multimedia data for a smooth display of them. Therefore, if the visual signal is delayed more than $D_\alpha$, we will retrieve the correct video frame from the buffer zone (i.e. take the next frame until synchronization status is judged); otherwise, if the visual signal is ahead by $D_\beta$, we will repeat the current frame until haptic-visual synchronization. Through this method, we are able to remove all asynchronization phenomena during haptic-visual delivery and playback.


\subsection{The overall method}
By summarizing Section 4.1-4.4, the detailed steps of our method are presented as follows.

\textbf{Step 1.} Initialization. Set a buffer zone at the receiving-end to cache haptic-visual data. Start the haptic-visual data delivery and playback. Go to Step 2.

\textbf{Step 2.} Key sample detection. Keep to detect the key samples of haptic signals with the method in Section 4.1. If a key sample is found, set the time interval as $T_h$ and go to Step 3.

\textbf{Step 3.} Key frame detection. Use the method in Section 4.2 to detect the corresponding key frames in the buffer and subsequent video of 1s. If a key frame is found, set the time interval as $T_v$ and go to Step 4; otherwise, the synchronization detection fails, go to Step 2.

\textbf{Step 4.} Asynchronization examination. If Eq.~(\ref{eq2}) of Section 4.3 is true, go to Step 2 to check the following signals; otherwise go to Step 5.

\textbf{Step 5.} Asynchronization removal. Adjust the haptic-visual streams with the method shown in Section 4.4. Go to Step 2 to check the following signals.

%

\section{Experimental Results}

To examine the effectiveness of proposed method, we implement it on the simulation platform shown in Section 3 and conduct both objective and subjective experiments. The frequencies of haptic and visual signals are set as 1000 Hz and 30 Hz, respectively. Due to the lack of haptic-visual synchronization method, we compare our model with the original case only. 

\subsection{Estimation accuracy of synchronization delay}

The proposed method utilizes the synchronization delay $T_v-T_h$ to determine whether asynchronization happens. Therefore, the estimation accuracy of synchronization delay is critical in our method. We design the following experiment to examine the accuracy.

Based on the simulation platform, we randomly captured 100 haptic-visual clips, with the length of each clip as 30s. In other words, there exist 30000 haptic samples and 900 video frames in each clip, and totally exist 3 million haptic samples and 90000 video frames). For each haptic-visual clip, we add a random delay on visual signals. The delay is in the range of [-330ms, 330ms] where the positive/negative values indicate visual signal is ahead/behind haptic signal. At the receiving-end, we employ our model to calculate the synchronization delay (namely \textit{\^{x}}) and compare it with the "actual" delay (namely $x$). The Mean Absolute Error (MAE) and Maximum Absolute Error (MaxAE) are utilized to be assessment metrics, in which MAE is calculated by:
\begin{equation}
	MAE=\dfrac{1}{M}{\sum_{i=1}^{M}}{\left| {\hat{x}_{i}}-x_{i} \right|},
\end{equation}
\noindent where M is the total number of samples.

The results are shown in Table~\ref{tab1}. From the table, the MAE and MaxAE values are 7.3 ms and 15 ms, respectively. It is noted that the haptic-visual synchronization is unperceivable in [-60 ms, 80 ms], where the ratio of MAE and MaxAE are only 5.2\% and 10.7\%, respectively. On the other hand, the frame length of each video frame is $\frac{1}{30}$Hz = 33.3 ms, which is also significantly larger than the MAE/MaxAE values. Therefore, the estimation accuracy could fulfill the requirement in the practical applications of haptic-visual system.

\begin{table}[htbp]
	\caption{The estimation accuracy of  $T_v-T_h$}
	\label{tab1}
	\begin{tabular}{ccc}
		\toprule
		Metrics&	MAE(ms)&	MaxAE  (ms)\\
		\midrule
		Results&	7.3	&15\\
		\bottomrule
	\end{tabular}
\end{table}


\begin{table*}[htbp]
	\caption{An example of random delay in the experiment}
	\label{tab+2}
	\begin{tabular}{ccccccccccccccccc}
		\toprule
	$d_n$&	7&	-7&	8&	-8&	8&	9&	-1&	0&	5& -8&	-8&	2	&0&	1&	-1	&-7		\\
	\toprule
	$t_n$&	19&	18	&95&	17&	56&	65&	82	&46&	69&
		96&	47&	86&	36	&99	&14&	55	\\
		\bottomrule
	\end{tabular}
\end{table*}

\begin{table}[htbp]
	\caption{Probabilities of synchronization with and without our method}
	\label{tab2}
	\begin{tabular}{ccc}
		\toprule
		&	Without our method	&With our method\\
		\midrule
	    Probabilities&	25.3\%   &89.2\%\\
		\bottomrule
	\end{tabular}
\end{table}

\begin{figure}[htbp]
	\centering
	\includegraphics[width=\linewidth]{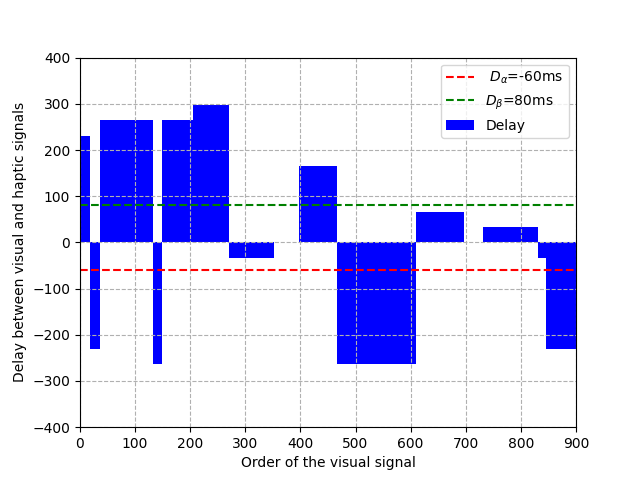}
	\caption{An example of random delay in the experiment}
	\Description{An example of random delay in the experiment}
	\label{picture+2}
\end{figure}

\subsection{Effectiveness of haptic-visual synchronization}

To evaluate the effectiveness of our synchronization detection and removal method, we examine it on the same dataset presented in Section 5.1.

At the sending end, after sending random video frames (in the range of [0, 100]), we add a random delay  (in the range of [-330ms, 330ms] and denoted as $t_n$) on it. We repeat the above process until all the frames in each chip (totally 100 clips) are sent. Considering that the proposed asynchronization removal method adjusts visual signal frame by frame, the interval of the above random delay is set the same as the frame interval of visual signal (i.e. 33ms). Therefore, the delay range of [-330ms, 330ms] is equivalent to delay random number (denoted as $d_n$) of video frames in the range of [-10, 10]. Taking a clip (900 frames) as example, the random numbers generated in the experiment are shown in Table~\ref{tab+2}. In the table, the values in the first column indicate that the visual signal is ahead of the haptic signal 7 $\times$ 33=231ms and the delay status lasts for 19 $\times$ 33=627ms. The above random delay in the experiment also intuitively shown in Fig.~\ref{picture+2}, in which the vertical axis indicates the delay between visual and haptic signals and the horizontal axis indicates the order of the visual signal. From the figure, the delays are random and representative to evaluate our method.

At the receiving end, we compare the probabilities of successful synchronization with and without our method. The results are presented in Table~\ref{tab2}. By using our model, the average probability of synchronization increases from 25.3\% to 89.2\%. It should be pointed out that our synchronization method is executed frame by frame. If the haptic-visual delay is larger than 1 frame, the signal is kept asynchronized during the synchronization process. That is the reason why there are still 10.8\% signals asynchronized in Table ~\ref{tab2}. Even at this scenario with severe fluctuations, our method still achieve a high probability of 89.2\%, which reveals the effectiveness and robustness of our method in haptic-visual synchronization. The utilization of our model guarantees the signal synchronization in most cases, thereby greatly improving the system performance of haptic-visual interaction.

\subsection{Subjective test on user experience}
Besides of objective evaluation, we also conduct subject test to evaluate the improvement of user experience by our model. As mentioned in Section 1, the signal asynchronization is a critical factor to influence user experience in haptic-visual interaction. Therefore, the improvement of user experience can be taken as circumstantial evidence of the effectiveness of our model.

We recruit 23 testees to participate this test, where all haptic-visual sequences are also the same to those in Section 5.1. To calculate the correlations, we introduce the delays that are evenly distributed from -10 to 10 frames ( that is, ranged from -333 ms to 333ms with the interval of 33.3ms) and occasionally utilize the proposed synchronization method at the receiving end. However, whether or not using the synchronization is unknown for all testees. As a result, a testee scores his/her experience based on the real feelings and experiences. All scores are between 0 and 10 and their averaged value, the Mean Opinion Score (MOS), represents the average perceptions of human users.

\begin{figure}[htbp]
	\centering
	\includegraphics[width=\linewidth]{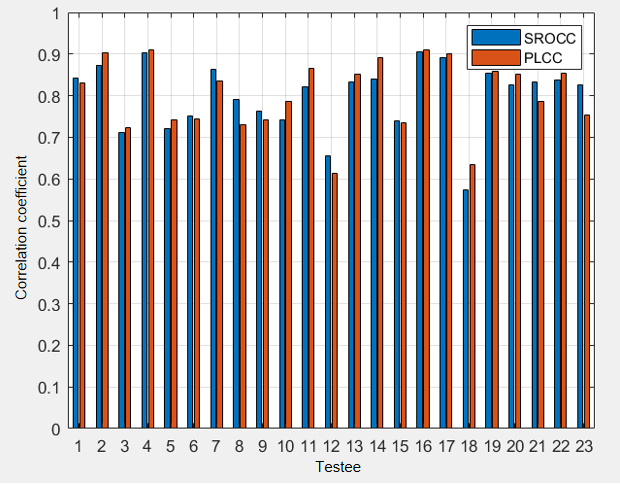}
	\caption{The correlations between each testee and the MOS}
	\Description{The correlations between each testee and the MOS}
	\label{picture7}
\end{figure}

\begin{figure}[htbp]
	\centering
	\includegraphics[width=\linewidth]{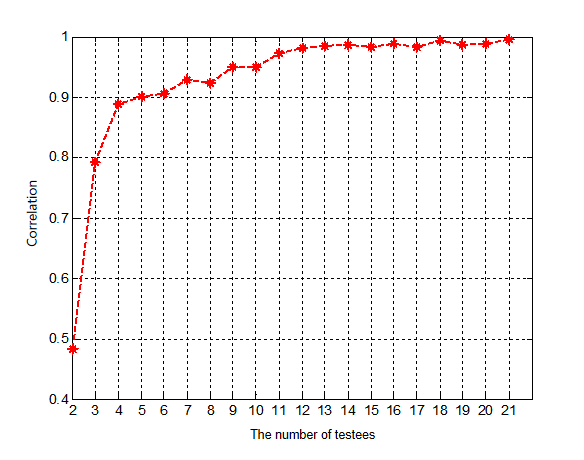}
	\caption{The data saturation validation in our test}
	\Description{The data saturation validation in our test}
	\label{picture8}
\end{figure}

The collected subjective test results are pre-processed to remove outliers based on the ITU subjective test regulations. We calculate the correlations, including Pearson Linear Correlation Coefficient (PLCC) and Spearman Rank-Order Correlation Coefficient (SROCC)~\cite{8212799}, between each testee's score and the MOS. The results are shown in Fig.~\ref{picture7}. According to ITU-R BT.500~\cite{123}, a testee's score is considered as an outlier if the correlation between his/her score and the MOS is less than 0.7. Therefore, from Fig.~\ref{picture7}, the 12th and 18th testees are considered as outliers and subsequently excluded in the final results.

\begin{figure}[htbp]
	\centering
	\includegraphics[width=\linewidth]{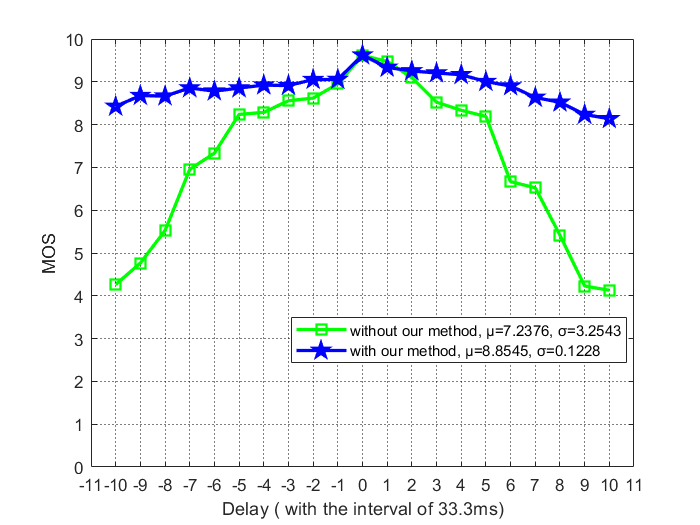}
	\caption{The subjective improvements with our method}
	\Description{The subjective improvements with our method}
	\label{picture9}
\end{figure}

The scores of the remaining 21 testees are further examined by data saturation validation~\cite{9148469}. Due to randomness of user scores, insufficient testees would lead to inaccurate MOS values. To check whether the testees are enough, the data saturation validation was proposed. For a subjective test with $K$ testees, it randomly selects $k=1, 2, …, K$ testees to calculate the correlation between their averaged score and the MOS. If the correlation value converges to 1 as $k$ increases, the testees are considered sufficient. In our test, this correlation value is very close to 1 with $k=13$ testees, as shown in Fig.~\ref{picture8}. Therefore, the remaining 21 testees are sufficient to represent the averaged opinion of human users.

Fig.~\ref{picture9} shows the MOS values under different delay settings. Two settings are compared: receiving end with and without our method. In the central part of curves (i.e. -33 ms ~ 66 ms), the delays are unperceivable to human users and thus the two settings achieve very similar MOS values. As the absolute value of delay gets larger, the difference between the two settings becomes more significant. In extreme cases (i.e. ±330ms), our synchronization method improves the MOS values by around 4, which shows the high capability of anti-interference under severe network conditions. On average, the MOS value is increased by 1.6169, with MOS variation decreased by 3.1315. This fact demonstrates the significant improvement of our synchronization method that is agreed by the majority of human users. In conclusion, the proposed method can guarantee the user experience in case of haptic-visual asynchronization.

\section{Conclusions}
In this paper, we exploited the haptic-visual correlations in haptic-aware interaction system. Based on the observations, we proposed a timestamp-independent synchronization method for haptic-visual signals, which consists of haptic signal analysis, learning-based vision analysis, perception-based thresholding and an overall method for asynchronization detection and removal. It should be pointed out that the referring of virtual hand (i.e. the ball) and target (i.e. the box) can be extended to more types of objects with retrained models. Therefore, our model is still applicable in more general scenarios. To our best knowledge, this is the very first attempt to design a haptic-aware multimedia synchronization model by considering the special characteristics of haptic interaction. It can also be utilized as a reference to design new synchronization models for emerging sensorial media such as olfaction signals. We envision a more widespread use of multiple sensorial media that benefits the immersive user experience in the foreseeable future.

\clearpage

\bibliographystyle{ACM-Reference-Format}
\bibliography{sample-base}


%
%
%
%
%
%
%

\end{document}